\documentstyle[twoside,fleqn,espcrc2]{article}

\title{
\vspace*{-35mm}
\rightline{
{\normalsize{NIKHEF 2000-010}}}
\vspace*{-2mm}
\rightline{\normalsize{June 2000}}
\vspace*{+20mm}
Deep-inelastic structure functions:
Reconstruction from Mellin moments}

\author{S. Moch{\thanks{Presented at 
{\it{Loops and Legs in Quantum Field Theory}}, 
9th - 14th April 2000, Bastei (Germany).}}
and J.A.M. Vermaseren\\
        NIKHEF Theory Group, 
        Kruislaan 409, 1098 SJ Amsterdam, The Netherlands}

\begin{document}
\pagestyle{empty}

\begin{abstract}
We summarize the method of calculating Mellin moments of deep-inelastic 
structure functions in perturbative QCD. 
We briefly discuss all steps to a complete analytical reconstruction 
of the perturbative corrrections in $x$-space.

\end{abstract}

\maketitle

\section{INTRODUCTION}

The precision measurement of structure functions and their 
scale evolution in deep inelastic scattering~\cite{Gross:1973rr,Gross:1974cs,Politzer:1973um} 
is of basic importance for testing perturbative QCD. 
By now, experimental data allows for an accurate determination 
of the parton distribution functions, which serve as input 
for numerous other hard scattering processes. 
Therefore, in particular with respect to the upcoming LHC experiments, 
it is highly desirable to further 
reduce the theoretical uncertainties on these very fundamental quantities.
This task requires the calculation of higher order perturbative QCD corrections.
In order to perform a consistent next-to-next-to-leading order 
analysis, which is expected to significantly reduce the theoretical 
uncertainties~\cite{vanNeerven:1999ca}, one has to calculate 
the still unknown anomalous dimensions of deep-inelastic structure functions 
to three loops. 

The method of calculating Mellin moments of deep-inelastic 
structure functions is directly connected with the early days of 
QCD~\cite{Gross:1973rr,Gross:1974cs,Politzer:1973um}.
Since then, it has been applied several times to determine
higher order perturbative corrections~\cite{Floratos:1977au,Floratos:1978gw,Floratos:1979ny,Gonzalez-Arroyo:1979df,Gonzalez-Arroyo:1980he,Lopez:1981dj,Kazakov:1988jk,Kazakov:1990jm,Larin:1991zw,Larin:1991tj,Larin:1997wd}, 
some of them related to sum rules. 
The method rests on the operator product expansion (OPE), which allows the calculation
of either all even or all odd integer moments $N$ of the anomalous dimensions 
and coefficient functions. 
This information by itself then uniquely determines the quantity under 
consideration for general $N$ in Mellin space or, equivalently 
its corresponding expression in $x$-space.

In the following, we will summarize the calculation 
of the anomalous dimensions and coefficient functions in terms of harmonic 
sums~\cite{Vermaseren:1998uu,Blumlein:1998if} 
and the subsequent analytical reconstruction based on an inverse Mellin transformation 
to harmonic polylogarithms~\cite{Remiddi:1999ew} in $x$-space. 
This procedure as detailed in~\cite{Moch:1999eb,Vermaseren:2000we} 
has recently been used~\cite{Moch:1999eb} to check the original calculation 
of the two-loop coefficient functions~\cite{vanNeerven:1991nn,Zijlstra:1991qc,Zijlstra:1992qd,Zijlstra:1992kj},
which was performed with conventional techniques.
Currently, work is in progress to obtain
the anomalous dimensions and coefficient functions at three loops by this method.

The present approach of analytical reconstruction nicely complements the use 
of Bernstein polynomials~\cite{Santiago:1999pr} for a direct fit of Mellin moments 
to experimental data. The latter method is particularly useful if only numerical results 
for a finite number of fixed moments are available, as it is presently the 
case~\cite{Larin:1997wd}.

\section{METHOD}

Deep-inelastic structure functions are defined by the commutator of two 
local electroweak currents $j(y)$ and $j(z)$, sandwiched between hadronic states 
and Fourier transformed into momentum space. 
In the Bjorken limit, for $Q^2 \to \infty$ and $x$ fixed, the OPE 
allows to express this current product in an asymptotic expansion 
around the lightcone $(y-z)^2 \sim 0$ into a series of local composite  
flavour non-singlet quark $O^{\alpha}$, singlet quark $O^{\rm q}$ 
and gluon operators $O^{\rm g}$ of leading twist and spin $N$. 
The same OPE also holds for the forward Compton amplitude $T$ 
of boson-hadron scattering. Standard perturbation theory applies to $T$, 
which can be written as a series in $1/x^N$ in the Euclidean region. 
Then, by means of the optical theorem one obtains a direct relation 
between the parameters of the OPE and the Mellin moments of the structure functions. 
For $F_{2}$ we can write
\begin{eqnarray}
\label{eq:F2mellin}
\displaystyle
F_{2}^N(Q^2)\!&=&\!     
\int\limits_0^1 dx\, x^{N-2} F_{2}(x,Q^2) \\
\!&=&\! 
\nonumber
\sum\limits_{j=\alpha,{\rm{q, g}}}
C_{2,j}^{N}\left(\frac{Q^2}{\mu^2},\alpha_s\right)
A_{{\rm{P}},N}^j\left(\mu^2\right)\, ,
\end{eqnarray}
and similar relations define $F_{3}^N$ and $F_{L}^N$.
Here $C_{2,j}^{N}$ denote the coefficient functions and 
$A_{{\rm{P}},N}^j$ the spin averaged hadronic matrix elements 
of the operators $O^{\alpha}$, $O^{\rm q}$ and $O^{\rm g}$.  

In the dispersive approach the derivation of eq.(\ref{eq:F2mellin}) 
uses symmetry properties of the Compton amplitude $T$ under exchange $x \to -x$.
As a consequence, eq.(\ref{eq:F2mellin}) is restricted to only either 
the even or the odd Mellin moments of $F_{2}$.
By analytic continuation on the other hand, all moments in the complex $N$-plane 
are fixed if the infinite set of either even or odd moments is known.
To be precise, for unpolarized electron-proton scattering eq.(\ref{eq:F2mellin}) 
determines the even moments of $F_{2}^{e{\rm P}}$, while for neutrino-proton scattering 
the odd moments of $F_{2}^{\nu{\rm{P}}-{\overline{\nu}{\rm{P}}}}$ 
and $F_{3}^{\nu{\rm{P}}+{\overline{\nu}{\rm{P}}}}$ 
and the even moments of $F_{2}^{\nu{\rm{P}}+{\overline{\nu}{\rm{P}}}}$ 
and $F_{3}^{\nu{\rm{P}}-{\overline{\nu}{\rm{P}}}}$ are fixed.

The coefficient functions and the renormalized operator matrix elements 
in eq.(\ref{eq:F2mellin}) both satisfy renormalization group equations.
Due to current conservation they are governed by the same anomalous dimensions, 
\begin{eqnarray}
\label{callanA}
{\lefteqn{
\sum_{k = \alpha,{\rm{q,g}}}
\Bigl[ \left\{ \mu^2 \frac{\partial}{\partial \mu^2} + \beta(\alpha_s(\mu^2))
 \frac{\partial}{\partial \alpha_s(\mu^2)} \right\} \delta_{jk} 
}} \\
& & + \gamma_{jk}(\alpha_s(\mu^2))
\Bigr]
A_{{\rm{P}},N}^j\left(\mu^2\right)
=   0 \, ,\nonumber \\
\label{callanC}
{\lefteqn{
\sum_{k = \alpha,{\rm{q,g}}}
\Bigl[ \left\{ \mu^2 \frac{\partial}{\partial \mu^2} + \beta(\alpha_s(\mu^2))
 \frac{\partial}{\partial \alpha_s(\mu^2)} \right\} \delta_{jk} 
}} \\
& & - \gamma_{jk}(\alpha_s(\mu^2))
\Bigr]
C^N_{2,k} \left(\frac{Q^2}{\mu^2},\alpha_s(\mu^2) \right) 
=   0 \, ,\nonumber
\end{eqnarray} 
where $j = \alpha,{\rm{q,g}}$, while $\beta$ and $\gamma_{jk}$
represent the QCD $\beta$-function and the anomalous dimensions.
Both are calculable order by order in $\alpha_s$ in perturbative QCD as well 
as the coefficient functions $C_{2,j}^{N}$.

The anomalous dimensions $\gamma_{jk}$ in eqs.(\ref{callanA}), (\ref{callanC}) 
determine the scale evolution of deep-inelastic structure functions and, 
dependent on the particular scattering process, 
one considers operator matrix elements corresponding 
to the quark flavour non-singlet and singlet distributions ${\rm{q}}_{{\rm ns}}^{\pm}$, 
${\rm{q}}_{{\rm ns}}^{\rm{V}}$ and ${\rm{q}}_{\rm{s}}$.  
Their scale evolution is governed by linear combinations,
\begin{eqnarray}
\label{splitting-functions-ns}
{\rm{q}}_{{\rm ns}}^{\pm} 
        & \longrightarrow& 
        \gamma_{{\rm{q}}{\rm{q}}}^{\rm{V}} \pm \gamma_{{\rm{q}}\bar{{\rm{q}}}}^{\rm{V}} \, \\
\label{splitting-functions-s-1}
{\rm{q}}_{\rm ns}^{\rm{V}} & \longrightarrow& 
        \gamma_{{\rm{q}}{\rm{q}}}^{\rm{V}} - \gamma_{{\rm{q}}\bar{{\rm{q}}}}^{\rm{V}} + n_f 
\left( \gamma_{{\rm{q}}{\rm{q}}}^{\rm{S}} - \gamma_{{\rm{q}}\bar{{\rm{q}}}}^{\rm{S}} \right)\, ,\\
\label{splitting-functions-s-2}
{\rm{q}}_{\rm s} & \longrightarrow&
        \gamma_{{\rm{q}}{\rm{q}}}^{\rm{V}} + \gamma_{{\rm{q}}\bar{{\rm{q}}}}^{\rm{V}} + 
                n_f \left(
        \gamma_{{\rm{q}}{\rm{q}}}^{\rm{S}} + \gamma_{{\rm{q}}\bar{{\rm{q}}}}^{\rm{S}} \right)\, ,
\end{eqnarray}
and the flavour singlet quark distribution ${\rm{q}}_{\rm{s}}$ 
mixes with the gluon distribution. 

The actual calculation of the anomalous dimensions $\gamma_{jk}$ and the 
coefficient functions $C_{2,j}^{N}$ is performed in dimensionally regularized 
perturbation theory with the external hadron states replaced by quark and gluon states 
of momentum $p$. 
By means of the OPE, the forward Compton amplitude for 
the scattering off a parton can be expressed in terms of 
the coefficient functions of eq.(\ref{eq:F2mellin}) and 
renormalized partonic operator matrix elements $A^{k}_{{\rm p},N}$. 
To be specific, we write for the flavour singlet case, 
\begin{eqnarray}
\lefteqn{
\label{TmunuPartonRenS}
T_{2,{\rm{p}}}(x,Q^2,\alpha_s,\epsilon) \,=} \\
& &
\sum_{N}
\sum_{j,k={\rm{q}},{\rm{g}}}  \left( \frac{1}{2x}\right)^N
C^{N}_{2,j}\left(\frac{Q^2}{\mu^2},\alpha_s,\epsilon\right)
\nonumber \\
& &
\times
Z^{jk}\left(\alpha_s,\frac{1}{\epsilon}\right)
A^{k}_{{\rm p},N}\left(\alpha_s,\mu^2,\epsilon\right)
+ O(p^2)\, , \nonumber
\end{eqnarray}
where ${{\rm{p}} = {\rm{q}},{\rm{g}}}$ and $O(p^2)$ denotes higher 
twist contributions. 
The $Z^{jk}$ represent the matrix of renormalization factors which contain
 all poles in $1/\epsilon$.
They relate bare to renormalized operators according to $O^{\rm{bare}} = Z \,O^{\rm{ren}}$ 
and provide the anomalous dimensions via,
\begin{eqnarray}
\label{anomdef}
\gamma &=& \displaystyle\left( \frac{d }{d \ln \mu^2 }
    Z\right) Z^{-1} \, .
\end{eqnarray}

The gauge invariant operators $O^{\rm{q}}$ and $O^{\rm{g}}$ in eq.(\ref{TmunuPartonRenS}) 
mix under renormalization with unphysical operators~\cite{Hamberg:1992qt,Matiounine:1998ky},
which are BRST variations of some operators or else disappear by use of the equations of motion. 
However, if one calculates quantities related to physical $S$-matrix elements with 
physical polarization and on-shell momenta such as in eq.(\ref{TmunuPartonRenS}), 
these unphysical operators vanish and have therefore been omitted 
in eq.(\ref{TmunuPartonRenS}).

The coefficient function $C_{2,j}^N$ and the renormalization factors $Z^{jk}$ 
are calculated from eq.(\ref{TmunuPartonRenS}) using the method of 
projection~\cite{Gorishnii:1983su,Gorishnii:1987gn}. 
This method relies on dimensional regularization and amounts to the application 
of the projection operator,
\begin{eqnarray}
\label{projectionoperator}
 {\cal P}_N \equiv
  \left. \Biggl[ \frac{q^{ \{\mu_1}\cdots q^{\mu_N \}}}{N !}
  \frac{\partial ^N}{\partial
p^{\mu_1} \cdots  \partial p^{\mu_N}} \Biggr] \right|_{p=0} \, ,
\end{eqnarray}
where $q^{ \{\mu_1}\cdots q^{\mu_N \}}$ is symmetrical 
and traceless part of the tensor $q^{\mu_1}\cdots q^{\mu_N }$.

It is obvious, that on the right hand side of eq.(\ref{TmunuPartonRenS})
the $N{\mbox{-th}}$ order differentiation in ${\cal P}_N$ projects on the precisely 
the $N{\mbox{-th}}$ moment, which is the coefficient of $1/(2x)^{N}$ being $(p\cdot\! q/Q^2)^N$. 
All higher powers of $p \cdot\! q/Q^2$ vanish after nullifying the momentum $p$. 
Furthermore, ${\cal P}_N$ does not act on the renormalization constants $Z^{jk}$ 
and the coefficient functions being only functions of $N$, $\alpha_s$ and $\epsilon$.
However, the nullification of the momentum $p$ in ${\cal P}_N$ effects 
the partonic matrix elements $A^{k}_{{\rm{p}},N}$.
There, it eliminates all diagrams containing loops, as they become massless tadpole 
diagrams, which are put to zero in dimensional regularization.
Only the tree level operator matrix elements $A^{p, {\rm{tree}}}_{{\rm{p}},N}$ survive. 

Additionally in the flavour singlet case this decouples the operator mixing. 
Thus, eq.(\ref{TmunuPartonRenS}) provides two independent identities 
to separately determine $Z^{\rm qq}$, $Z^{\rm gq}$ 
and $Z^{\rm qg}$, $Z^{\rm gg}$.
We should be aware however, that an $l$-loop calculation 
eq.(\ref{TmunuPartonRenS}) alone does not suffice to give the full information 
about the renormalization constants $Z^{\rm gq} $ and $Z^{\rm gg}$. 
They are determined only up to $(l-1)$-loops. 
This limitation is due to the gluonic coefficient function $C^N_{2,{\rm{g}}}$ 
being zero at tree level, as photons couple directly to quarks only. 
To resolve this situation and to extract the anomalous dimension 
$\gamma_{\rm{gq}}$ and $\gamma_{\rm{gg}}$ to $l$-loops we also calculate 
Green's functions in which the photon is replaced by an external 
scalar particle $\phi$ that couples directly only to gluons~\cite{Larin:1997wd}.
These Green's functions are obtained from a gauge invariant interaction term 
$\phi F_{\mu\nu}^{a}F^{\mu\nu}_a$ in the QCD Lagrangian, 
$F_{\mu\nu}^{a}$ being the QCD field strength.
They can be expressed in partonic invariants $T_{\phi,\rm p}$ and satisfy 
an OPE similar to eqs.(\ref{eq:F2mellin}),(\ref{TmunuPartonRenS}) 
with the same singlet operators $O^{\rm q}$ and $O^{\rm g}$ but with different 
coefficient functions $C^N_{\phi,{\rm p}}$, where the gluonic coefficient 
function $C^N_{\phi,{\rm{g}}}$ starts now already tree level. 
Thus the invariants $T_{\phi,\rm p}$ provide us with the necessary 
renormalization constants $Z^{\rm gq} $ and $Z^{\rm gg}$ of the singlet operators 
to $l$-loops and complete the determination of all anomalous dimensions 
in the flavour singlet case to the desired order.

Use of the method of projection to calculate all Feynman diagrams 
contributing to the forward partonic Compton amplitude, that is to say 
the application of the projection operator ${\cal P}_N$ to the integrands 
on the left hand side of eq.(\ref{TmunuPartonRenS}) proceeds as follows.
All ultraviolet divergences as well as those of infrared origin 
as the momentum $p\rightarrow 0$ are dimensionally regularized.
At the same time, the nullification of $p$ drastically simplifies the topological structure 
of the corresponding graphs, which is reduced to the type of self-energy diagrams. 
Although the latter appear with symbolic powers of scalar products in the numerator 
and denominator, they are by far easier to calculate. 

In practice~\cite{Moch:1999eb,Vermaseren:2000we} 
this is performed by means of integration-by-parts identities and 
scaling identities for the $N{\mbox{-th}}$ Mellin moment of the diagram to be calculated.
Combined these identities give rise to difference equations, which relate the 
$N{\mbox{-th}}$ Mellin moment of the diagram under consideration to lower Mellin moments 
and diagrams of a simpler topology. 
Thereby they introduce a natural hierarchy in the set of diagrams. 
All difference equations can be summed recursively and the answer 
is expressible in terms of a basis of harmonic sums~\cite{Vermaseren:1998uu,Blumlein:1998if} 
of a given weight.
The whole approach exploits powerful summation algorithms~\cite{Vermaseren:1998uu} 
for a large class of single parameter sums which are reducible to the basis of harmonic sums.

As a demonstration of the method discussed thus far and as a check 
on~\cite{vanNeerven:1991nn,Zijlstra:1991qc,Zijlstra:1992qd,Zijlstra:1992kj},
we have performed a recalculation 
of the two-loop coefficient functions.
The contributing Feynman diagrams have been generated with 
QGRAF~\cite{Nogueira:1991ex} and all recursion relations for the evaluation 
of the individual topologies have been programmed in FORM~\cite {FORM}. 
The nested sums which one encounters in this way are solved with 
the SUMMER algorithm~\cite{Vermaseren:1998uu} in terms of the basis of harmonic sums.
In addition, there is the possibility to perform checks at all stages 
of the calculation by means of the standard MINCER routine~\cite{Larin:1991fz}.
This last feature is very important from a practical point of view as 
the debugging is greatly simplified.

To choose a specific example, we consider the case of initial state quarks 
in $D=4-2\epsilon$ dimensions and calculate with the help of 
eqs.(\ref{TmunuPartonRenS}),(\ref{projectionoperator}) 
the projected invariant ${\cal P}_N T_{2,{\rm{q}}}$.
After performing the  coupling constant renormalization one is only left with 
collinear singularities associated with the initial state quarks. 
Expanding ${\cal P}_N T_{2,{\rm{q}}}$ in $\alpha_s$ 
and $\epsilon$ gives at leading order
\begin{eqnarray}
\label{F2-0}
T^{(0)}_{2,{\rm{q}}} = 1\, .
\end{eqnarray}
At first order in $\alpha_s$, we have to expand up to order $\epsilon$ and find,
\begin{eqnarray}
\label{F2-1}
{\lefteqn{
T^{(1)}_{2,{\rm{q}}} \, = }}\\
& &\frac{\alpha_s}{4 \pi}\, S_\epsilon\, \left( \frac{\mu^2}{Q^2} \right)^{\epsilon}
        \left[ \frac{1}{\epsilon} \gamma^{(0)}_{\rm{qq}} + 
                c^{(1)}_{2,{\rm{q}}}
                        + \epsilon a^{(1)}_{2,{\rm{q}}}
                \right] \, ,\nonumber
\end{eqnarray}
where the factor $S_\epsilon = \exp( \epsilon\{\ln(4\pi) - \gamma_{\rm{E}}\} )$
and $\gamma^{(n)}_{\rm qq}$, $c^{(n)}_{2, \rm q}$ are the coefficients of 
$(\alpha_s/4 \pi)^n$ in the expansion of the anomalous dimension $\gamma_{\rm qq}$ and 
the quark coefficient function $C^N_{2,{\rm{q}}}$.

At second order in $\alpha_s$, we need to split up the contributions into flavour
non-singlet and singlet parts,
\begin{eqnarray}
T^{(2),\rm{s}}_{2,{\rm{q}}} \,=\, T^{(2),\rm{ns},+}_{2,{\rm{q}}} 
                + T^{(2),\rm{ps}}_{2,{\rm{q}}}\, .
\end{eqnarray}
Allowing for electroweak interactions, one can consider 
in the non-singlet case the structure functions of different physical processes,
$F_{2}^{\nu{\rm{P}}\pm{\overline{\nu}{\rm{P}}}}$, which implies a distinction 
of even and odd moments, $T^{(2),\rm{ns},+}_{2,{\rm{q}}}$ and $T^{(2),\rm{ns},-}_{2,{\rm{q}}}$ 
to be precise. 
We have 
\begin{eqnarray}
\label{F2-NS-2}
{\lefteqn{
T^{(2),\rm{ns},\pm}_{2,{\rm{q}}} \,=\, 
\left( \frac{\alpha_s}{4 \pi} \right)^2 S_\epsilon^2\,
\left( \frac{\mu^2}{Q^2} \right)^{2\epsilon} }} \\
& &\times
\left[ \frac{1}{\epsilon^2} \frac{1}{2} 
\left\{ \! \Bigl(\gamma^{(0)}_{\rm{qq}}\Bigr)^2 - \beta_0 \gamma^{(0)}_{\rm{qq}} \right\} 
        + \frac{1}{\epsilon} \left\{ 
          \frac{1}{2} \gamma^{(1),\pm,\rm{V}}_{\rm{qq}} 
\right.\right. 
\nonumber\\
& &
\left.\left.
\hspace*{10pt}
                + \gamma^{(0)}_{\rm{qq}} c^{(1)}_{2,{\rm{q}}} 
                \right\} 
+ c^{(2),\rm{ns},\pm}_{2,{\rm{q}}} + \gamma^{(0)}_{\rm{qq}} a^{(1)}_{2,{\rm{q}}} 
                \right] , \nonumber
\end{eqnarray}
where $\gamma_{{\rm{q}}{\rm{q}}}^{\pm,\rm{V}}$ denotes the combinations 
$\gamma_{{\rm{q}}{\rm{q}}}^{\rm{V}} \pm \gamma_{{\rm{q}}\bar{{\rm{q}}}}^{\rm{V}}$ 
as given in eqs.(\ref{splitting-functions-ns})--(\ref{splitting-functions-s-2}).

The pure-singlet contributions on the other hand give at second order in $\alpha_s$, 
\begin{eqnarray}
\label{F2-PS-2}
{\lefteqn{
T^{(2),\rm{ps}}_{2,{\rm{q}}} \,=\,
 n_f\, \left(\frac{\alpha_s}{4 \pi} \right)^2 S_\epsilon^2\,
\left( \frac{\mu^2}{Q^2} \right)^{2\epsilon}}} \\
& &\times
 \left[ \frac{1}{\epsilon^2} 
\frac{1}{2} \gamma^{(0)}_{\rm{qg}} \gamma^{(0)}_{\rm{gq}} 
        + \frac{1}{\epsilon} \left\{ \frac{1}{2} \gamma^{(1),+,\rm{S}}_{\rm{qq}} 
                + \gamma^{(0)}_{\rm{gq}} c^{(1)}_{2,{\rm{g}}} \right\} 
\right. \nonumber\\
& & 
\left.\hspace*{10pt}
+ c^{(2),\rm{ps}}_{2,{\rm{q}}} + \gamma^{(0)}_{\rm{gq}} a^{(1)}_{2,{\rm{g}}}
                \right] , \nonumber 
\end{eqnarray}
and $\gamma_{{\rm{q}}{\rm{q}}}^{\pm,\rm{S}}$ represents the combinations 
$\gamma_{{\rm{q}}{\rm{q}}}^{\rm{S}} \pm \gamma_{{\rm{q}}\bar{{\rm{q}}}}^{\rm{S}}$ 
of eqs.(\ref{splitting-functions-s-1}),(\ref{splitting-functions-s-2}).

With eqs.(\ref{F2-0}),(\ref{F2-1}) and (\ref{F2-NS-2}),(\ref{F2-PS-2}) at hand, we have 
at this point calculated all anomalous dimension and coefficient functions up to two loops 
in Mellin space as analytical functions of the moment $N$. 
Depending on the physical process under consideration, the quantities of interest 
are defined for either even or odd $N$. 
In a following step, we perform an inverse Mellin transformation to 
obtain the corresponding expressions in $x$-space.
As discussed in~\cite{Remiddi:1999ew,Moch:1999eb,Vermaseren:2000we}, 
we can do this step analytically as there exists a one-to-one 
mapping~\cite{Remiddi:1999ew,Vermaseren:2000we} of harmonic sums of weight $w+1$ 
to the set of harmonic polylogarithms of weight $w$ multiplied by $1/(1\pm x)$.
The latter one is the natural class of functions for calculations of deep-inelastic 
structure functions. 
Since the resulting expressions in $x$-space are unique, we may subsequently 
execute another Mellin transformation to go back to $N$-space, which then serves 
as an analytical continuation to all non-negative integer values in $N$.

In particular, we are then ready to determine individually all anomalous dimensions 
which control the scale evolution of the structure functions in the flavour non-singlet case 
as entering in eqs.(\ref{splitting-functions-ns})--(\ref{splitting-functions-s-2}).
This is easily done now, as we obtain from the even moments of $F_{2}^{e{\rm P}}$ 
the sums $\gamma_{{\rm{q}}{\rm{q}}}^{+,\rm{V}}$ and $\gamma_{{\rm{q}}{\rm{q}}}^{+,\rm{S}}$.
The odd moments of $F_{2}^{\nu{\rm{P}}-{\overline{\nu}{\rm{P}}}}$ on the other hand
determine the differences 
$\gamma_{{\rm{q}}{\rm{q}}}^{-,\rm{V}}$ and $\gamma_{{\rm{q}}{\rm{q}}}^{-,\rm{S}}$, 
the latter being zero up to two loops.
As emphasized above, the analytical continuation provides us with expressions valid 
for all non-negative integer values of $N$, such that at this point we can simply 
take the sums and the differences of $\gamma_{{\rm{q}}{\rm{q}}}^{+,\rm{V,S}}$ and 
$\gamma_{{\rm{q}}{\rm{q}}}^{-,\rm{V,S}}$ to obtain all quantities 
$\gamma_{{\rm{qq}}}^{\rm{V}}, \gamma_{{\rm{q}}\bar{{\rm{q}}}}^{\rm{V}},
\gamma_{{\rm{qq}}}^{\rm{S}}$ and $\gamma_{{\rm{q}}\bar{{\rm{q}}}}^{\rm{S}}$ 
of eqs.(\ref{splitting-functions-ns})--(\ref{splitting-functions-s-2}).

Similar reconstructions in Mellin space and $x$-space may be carried out for 
the two loop coefficient functions in eqs.(\ref{F2-NS-2}),(\ref{F2-PS-2}). 
To summarize, we find complete agreement with the published results for the 
anomalous dimensions~\cite{Floratos:1977au,Floratos:1978gw,Floratos:1979ny,Gonzalez-Arroyo:1979df,Gonzalez-Arroyo:1980he,Lopez:1981dj,Curci:1980uw,Furmanski:1980cm}
and for the coefficient functions~\cite{vanNeerven:1991nn,Zijlstra:1991qc,Zijlstra:1992qd,Zijlstra:1992kj}.

\section{CONCLUSION}

The calculation of higher order pertubative corrections to deep-inelastic 
structure functions is vitally important to improve quantitative predictions 
for many hard scattering processes in QCD.

In the past, this has been done either in $x$-space or in Mellin space. 
Using the method of projection, which enables us to obtain anomalous dimensions 
and coefficient functions at the same time, we perform the calculation in Mellin space.
Due to new insight into the mathematical properties of harmonic sums 
and their interplay with harmonic polylogarithms, we are first of all able to solve all 
nested sums in Mellin space. 
Secondly we can reconstruct the complete analytical expressions 
of the results in $x$-space by means of an inverse Mellin transformation.
Thereby we exploit all advantages of working in Mellin space, such as the use of 
difference equations in the Mellin moment $N$ to solve the resulting loop integrals 
and the possibility for independent checking with the MINCER routine.

Finally, we would like to remark, that the method presented allows 
for a direct application to the calculation of the three loop anomalous dimensions, 
as well as a generalization to polarized deep inelastic scattering~\cite{MVinprep}.


\begin{thebibliography}{10}

\bibitem{Gross:1973rr}
D.~J. Gross and F.~Wilczek,
\newblock Phys. Rev. {\bf D8}, 3633 (1973).

\bibitem{Gross:1974cs}
D.~J. Gross and F.~Wilczek,
\newblock Phys. Rev. {\bf D9}, 980 (1974).

\bibitem{Politzer:1973um}
H.~D. Politzer,
\newblock Phys. Rev. Lett. {\bf 30}, 1346 (1973).

\bibitem{vanNeerven:1999ca}
W.~L. van Neerven and A.~Vogt,
\newblock Nucl. Phys. {\bf B568}, 263 (2000), hep-ph/9907472.

\bibitem{Floratos:1977au}
E.~G. Floratos, D.~A. Ross, and C.~T. Sachrajda,
\newblock Nucl. Phys. {\bf B129}, 66 (1977).

\bibitem{Floratos:1978gw}
E.~G. Floratos, D.~A. Ross, and C.~T. Sachrajda,
\newblock Nucl. Phys. {\bf B139}, 545 (1978).

\bibitem{Floratos:1979ny}
E.~G. Floratos, D.~A. Ross, and C.~T. Sachrajda,
\newblock Nucl. Phys. {\bf B152}, 493 (1979).

\bibitem{Gonzalez-Arroyo:1979df}
A.~Gonzalez-Arroyo, C.~Lopez, and F.~J. Yndurain,
\newblock Nucl. Phys. {\bf B153}, 161 (1979).

\bibitem{Gonzalez-Arroyo:1980he}
A.~Gonzalez-Arroyo and C.~Lopez,
\newblock Nucl. Phys. {\bf B166}, 429 (1980).

\bibitem{Lopez:1981dj}
C.~Lopez and F.~J. Yndurain,
\newblock Nucl. Phys. {\bf B183}, 157 (1981).

\bibitem{Kazakov:1988jk}
D.~I. Kazakov and A.~V. Kotikov,
\newblock Nucl. Phys. {\bf B307}, 721 (1988).

\bibitem{Kazakov:1990jm}
D.~I. Kazakov and A.~V. Kotikov,
\newblock Nucl. Phys. {\bf B345}, 299 (1990),
\newblock Erratum.

\bibitem{Larin:1991zw}
S.~A. Larin, F.~V. Tkachev, and J.~A.~M. Vermaseren,
\newblock Phys. Rev. Lett. {\bf 66}, 862 (1991).

\bibitem{Larin:1991tj}
S.~A. Larin and J.~A.~M. Vermaseren,
\newblock Phys. Lett. {\bf B259}, 345 (1991).

\bibitem{Larin:1997wd}
S.~A. Larin, P.~Nogueira, T.~van Ritbergen, and J.~A.~M. Vermaseren,
\newblock Nucl. Phys. {\bf B492}, 338 (1997), hep-ph/9605317.

\bibitem{Vermaseren:1998uu}
J.~A.~M. Vermaseren,
\newblock Int. J. Mod. Phys. {\bf A14}, 2037 (1999), hep-ph/9806280.

\bibitem{Blumlein:1998if}
J.~Blumlein and S.~Kurth,
\newblock Phys. Rev. {\bf D60}, 014018 (1999), hep-ph/9810241.

\bibitem{Remiddi:1999ew}
E.~Remiddi and J.~A.~M. Vermaseren,
\newblock (1999), hep-ph/9905237.

\bibitem{Moch:1999eb}
S.~Moch and J.~A.~M. Vermaseren,
\newblock Nucl. Phys. {\bf B573}, 853 (2000), hep-ph/9912355.

\bibitem{Vermaseren:2000we}
J.~A.~M. Vermaseren and S.~Moch,
\newblock (2000), hep-ph/0004235,
\newblock these proceedings.

\bibitem{vanNeerven:1991nn}
W.~L. van Neerven and E.~B. Zijlstra,
\newblock Phys. Lett. {\bf B272}, 127 (1991).

\bibitem{Zijlstra:1991qc}
E.~B. Zijlstra and W.~L. van Neerven,
\newblock Phys. Lett. {\bf B273}, 476 (1991).

\bibitem{Zijlstra:1992qd}
E.~B. Zijlstra and W.~L. van Neerven,
\newblock Nucl. Phys. {\bf B383}, 525 (1992).

\bibitem{Zijlstra:1992kj}
E.~B. Zijlstra and W.~L. van Neerven,
\newblock Phys. Lett. {\bf B297}, 377 (1992).

\bibitem{Santiago:1999pr}
J.~Santiago and F.~J. Yndurain,
\newblock Nucl. Phys. {\bf B563}, 45 (1999), hep-ph/9904344.

\bibitem{Hamberg:1992qt}
R.~Hamberg and W.~L. van Neerven,
\newblock Nucl. Phys. {\bf B379}, 143 (1992).

\bibitem{Matiounine:1998ky}
Y.~Matiounine, J.~Smith, and W.~L. van Neerven,
\newblock Phys. Rev. {\bf D57}, 6701 (1998), hep-ph/9801224.

\bibitem{Gorishnii:1983su}
S.~G. Gorishnii, S.~A. Larin, and F.~V. Tkachev,
\newblock Phys. Lett. {\bf 124B}, 217 (1983).

\bibitem{Gorishnii:1987gn}
S.~G. Gorishnii and S.~A. Larin,
\newblock Nucl. Phys. {\bf B283}, 452 (1987).

\bibitem{Nogueira:1991ex}
P.~Nogueira,
\newblock J. Comput. Phys. {\bf 105}, 279 (1993).

\bibitem{FORM}
J.~A.~M. Vermaseren,
\newblock {\em Symbolic Manipulation with FORM} (Computer Algebra Nederland,
  Kruislaan 413, 1098 SJ Amsterdam, 1991),
\newblock ISBN 90-74116-01-9.

\bibitem{Larin:1991fz}
S.~A. Larin, F.~V. Tkachev, and J.~A.~M. Vermaseren,
\newblock NIKHEF-H-91-18.

\bibitem{Curci:1980uw}
G.~Curci, W.~Furmanski, and R.~Petronzio,
\newblock Nucl. Phys. {\bf B175}, 27 (1980).

\bibitem{Furmanski:1980cm}
W.~Furmanski and R.~Petronzio,
\newblock Phys. Lett. {\bf 97B}, 437 (1980).

\bibitem{MVinprep}
S.~Moch and J.~A.~M. Vermaseren,
\newblock in preparation.

\end{thebibliography}
\end{document}